\documentclass[structabstract]{aa}  
\usepackage{graphicx}
\usepackage{txfonts}
\usepackage{natbib}
\usepackage{color}
\usepackage{multirow}

\begin{document}

	\title{Extended envelopes around Galactic Cepheids}
	\titlerunning{Extended envelopes around Cepheids}

   	\subtitle{IV.~T~Monocerotis and X~Sagittarii from mid-infrared interferometry with VLTI/MIDI\thanks{Based on observations made with ESO telescopes at Paranal observatory under program ID 082.D-0066}}

	\author{ A.~Gallenne\inst{1} \and
				A.~M\'erand\inst{2} \and  
  				P.~Kervella\inst{3} \and
  				O.~Chesneau\inst{4} \and
  				J.~Breitfelder\inst{2,3}\and
				W.~Gieren\inst{1} 		
  				}
  				
  	\authorrunning{A. Gallenne et al.}

\institute{Universidad de Concepci\'on, Departamento de Astronom\'ia, Casilla 160-C, Concepci\'on, Chile
	\and LESIA, Observatoire de Paris, CNRS UMR 8109, UPMC, Universit\'e
  Paris Diderot, 5 Place Jules Janssen, F-92195 Meudon, France
  	\and European Southern Observatory, Alonso de C\'ordova 3107,
  Casilla 19001, Santiago 19, Chile
  \and Nice Sophia-Antipolis University, CNRS UMR 6525, Observatoire de la C\^ote d’Azur, BP 4229, 06304, Nice Cedex 4, France}%\\
  
  \offprints{A. Gallenne} \mail{agallenne@astro-udec.cl}

   \date{Received July 11, 2013; accepted August 30, 2013}

% \abstract{}{}{}{}{} 
% 5 {} token are mandatory
 
  \abstract
  % context heading (optional)
  % {} leave it empty if necessary  
  {}
  % aims heading (mandatory)
   {We study the close environment of nearby Cepheids using high spatial resolution observations in the mid-infrared with the VLTI/MIDI instrument, a two-beam interferometric recombiner.}
  % methods heading (mandatory)
   {We obtained spectra and visibilities for the classical Cepheids X~Sgr and T~Mon. We fitted the MIDI measurements, supplemented by $B, V, J, H, K$ literature photometry, with the numerical transfer code \texttt{DUSTY} to determine the dust shell parameters. We used a typical dust composition for circumstellar environments.}
  % results heading (mandatory)
   {We detect an extended dusty environment in the spectra and visibilities for both stars, although T~Mon might suffer from thermal background contamination. We attribute this to the presence of a circumstellar envelope (CSE) surrounding the Cepheids. This is optically thin for X~Sgr ($\tau_\mathrm{0.55\mathrm{\mu m}} = 0.008$), while it appears to be thicker for T~Mon ($\tau_\mathrm{0.55\mathrm{\mu m}} = 0.15$). They are located at about 15--20 stellar radii. Following our previous work, we derived a likely period-excess relation in the VISIR PAH1 filter, $ f_\mathrm{8.6\,\mu m}$[\%]$ = 0.81(\pm0.04)P$[day]. We argue that the impact of CSEs on the mid-IR period--luminosity (P--L) relation cannot be negligible because they can bias the Cepheid brightness by up to about 30\,\%. For the $K$-band P--L relation, the CSE contribution seems to be lower ($< 5$\,\%), but the sample needs to be enlarged to firmly conclude that the impact of the CSEs is negligible in this band.}
  % conclusions heading (optional), leave it empty if necessary 
   {}

 
 \maketitle

%
%________________________________________________________________

\section{Introduction}

A significant fraction of classical Cepheids exhibits an infrared excess, that is probably caused by a circumstellar envelope (CSE). The discovery of the first CSE around the Cepheid $\ell$~Car made use of near- and mid-infrared interferometric observations \citep{Kervella_2006_03_0}. Similar detections were subsequently reported for other Cepheids \citep{Merand_2007_08_0,Merand_2006_07_0,Barmby_2011_11_0,Gallenne_2011_11_0}, leading to the hypothesis that maybe all Cepheids are surrounded by a CSE. %Moreover, it is known from other techniques that they experience significant mass loss, possibly related to the existence of these CSEs. 

These envelopes are interesting from several aspects. Firstly, they might be related to past or ongoing stellar mass loss and might be used to trace the Cepheid evolution history. Secondly, their presence might induce a bias to distance determinations made with Baade-Wesselink methods and bias the calibration of the IR period--luminosity (P--L) relation. Our previous works \citep{Gallenne_2011_11_0,Merand_2007_08_0,Merand_2006_07_0,Kervella_2006_03_0} showed that these CSEs have an angular size of a few stellar radii and a flux contribution to the photosphere ranging from a few percent to several tens of percent. While in the near-IR the CSE flux emission might be negligible compared with the photospheric continuum, this is not the case in the mid- and far-IR, where the CSE emission dominates \citep{Gallenne_2011_11_0,Kervella_2009_05_0}.

Interestingly, a correlation starts to appear between the pulsation period and the CSE brightness in the near- and mid-IR bands: long-period Cepheids seem to show relatively brighter CSEs than short-period Cepheids, indicating that the mass-loss mechanism could be linked to stellar pulsation \citep{Gallenne_2011_11_0,Merand_2007_08_0}. Cepheids with long periods have higher masses and larger radii, therefore if we assume that the CSE IR brightness is an indicator of the mass-loss rate, this would mean that heavier stars experience higher mass-loss rates. This behavior could be explained by the stronger velocity fields in longer-period Cepheids and shock waves at certain pulsation phases \citep{Nardetto_2008_10_0,Nardetto_2006_07_0}. Studying this correlation between the pulsation period and the IR excess is vital for calibrating relations between the Cepheid' fundamental parameters with respect to their pulsation periods. If CSEs substantially influence the observational estimation of these fundamental parameters (luminosity, mass, radius, etc.), this a correlation will lead to a biased calibration. It is therefore essential to continue studying and characterizing these CSEs and to increase the statistical sample to confirm their properties.

We present new spatially resolved VLTI/MIDI interferometric observations of the classical Cepheids \object{X~Sgr} (HD~161592, $P = 7.01$\,days) and \object{T~Mon} (HD~44990, $P = 27.02$\,days). The paper is organized as follows. Observations and data reduction procedures are presented in Sect.~\ref{section__observation}. The data modeling and results are reported in Sect.~\ref{section__cse_modeling}. In Sect.~\ref{section__period_excess_relation} we address the possible relation between the pulsation period and the IR excess. We then discuss our results in Sect.~\ref{section__discussion} and conclude in Sect.~\ref{section__conclusion}.

\section{VLTI/MIDI observations}
\label{section__observation}

\subsection{Observations}

The observations were carried out in 2008 and 2009 with the VLT Unit Telescopes and the MIDI instrument \citep{Leinert_2003__0}. MIDI combines the coherent light coming from two telescopes in the $N$ band ($\lambda = 8-13\,\mu$m) and provides the spectrum and spectrally dispersed fringes with two possible spectral resolutions ($R = \Delta \lambda / \lambda = 30, 230$). For the observations presented here, we used the prism that provides the lowest spectral resolution. During the observations, the secondary mirrors of the two Unit Telescopes (UT1-UT4) were chopped with a frequency of 2 Hz to properly sample the sky background. MIDI has two photometric calibration modes: HIGH\_SENS, in which the flux is measured separately after the interferometric observations, and SCI\_PHOT, in which the photometry is measured simultaneously with the interferences fringes. These reported observations were obtained in HIGH\_SENS mode because of a relatively low thermal IR brightness of our Cepheids.

To remove the instrumental and atmospheric signatures, calibrators of known intrinsic visibility were observed immediately before or after the Cepheid. They were chosen from the \citet{Cohen_1999_04_0} catalog, and are almost unresolved at our projected baselines ($V > 95$\,\%, except for HD~169916, for which $V = 87$\,\%). The systematic uncertainty associated with their a priori angular diameter error bars is negligible compared with the typical precision of the MIDI visibilities (10--15\,\%). The uniform-disk angular diameters for the calibrators as well as the corresponding IRAS 12\,$\mu$m flux and the spectral type are given in Table~\ref{table__calibrators}.

The log of the MIDI observations is given in Table~\ref{table__journal}. Observations \#1, \#2, and \#5--\#10 were not used because of low interferometric or/and photometric flux, possibly due to a temporary burst of very bad seeing or thin cirrus clouds.

\begin{table}[!t]
\centering
\caption{Properties of our calibrator stars.}
\begin{tabular}{ccccc}
	\hline\hline
	HD & $\theta_\mathrm{UD}$	&  $f_\mathrm{W3}$	& 	$f_\mathrm{Cohen}$	&	Sp. Type	\\
			&		(mas)							& (Jy)	&	(Jy)		&					\\							
	\hline
	49293   & $1.91 \pm 0.02$ &	$4.3 \pm 0.1$ 	  &  $4.7 \pm 0.1$	&	K0IIIa \\
	48433   & $2.07 \pm 0.03$ &	$6.5 \pm 0.1$  	   & $5.5 \pm 0.1$		&	K0.5III \\
	168592 & $2.66 \pm 0.05$ &	$8.3 \pm 0.1$	& 	$7.4 \pm 0.1$	&	K4-5III \\
	169916 & $4.24 \pm 0.05$ &	$25.9 \pm 0.4$   & $21.1 \pm 0.1$		&	K1IIIb \\
	\hline
\end{tabular}
\tablefoot{$\theta_\mathrm{UD}$ is the uniform disk angular diameter, $f_\mathrm{W3}$ denotes the WISE flux \citep{Wright_2010_12_0} in the W3 filter at $11.6\,\mathrm{\mu m}$, while $f_\mathrm{Cohen}$ stands for the $12\,\mathrm{\mu m}$ monochromatic  flux from \citet{Cohen_1999_04_0}.}
\label{table__calibrators}
\end{table}

\begin{table}[!t]
\centering
\caption{Log of the observations.}
\begin{tabular}{ccccccc}
	\hline\hline
	\# &    MJD    		& $\phi$ & Target 					   & $B_\mathrm{p}$		   & PA                &  AM  \\
		 &					 &			   &								 & (m)							  & ($\degr$)	   &		\\
	\hline
	1   & 54~813.26 &  0.12  & T~Mon  					    &			129.6       	 &     63.7         & 1.20 \\
	2   & 54~813.27 &  0.12  & T~Mon  					    &			130.0       	 &     63.5         & 1.21 \\
	3	& 54~813.27 &			 & \object{HD~49293}   &  		    130.0			 &	   63.4			& 1.15 \\
	4	& 54~813.29 &	0.12  & T~Mon					    &			130.0			 &	   62.6			& 1.26 \\
	5	& 54~813.30 &	0.12  & T~Mon					    &			129.6			 &	   62.2			& 1.29 \\
	6	& 54~813.31 &			 & \object{HD~48433}   &			130.0			 &	   61.6			& 1.40 \\
	\hline
	7	& 54~842.10 & 0.18   & T~Mon					    &		   108.7			 &	   62.6		    & 1.26 \\
	8	& 54~842.11 &			  &	HD~49293			     &		    110.1			  &		59.7		 & 1.21 \\
	9	& 54~842.13 & 0.18   & T~Mon					    &		   118.8			 &	   64.4			& 1.19 \\
	10	& 54~842.14 &		     & HD~49293			     &	        120.8			  &		62.1		 & 1.14 \\
	11	& 54~900.07 & 0.33  & T~Mon					    &		   128.5			 & 	   61.4			& 1.33 \\
	12 & 54~900.09 &			 & HD~48433   &		  128.8				&     59.9		   & 1.49 \\
	13 & 54~905.39 &			 & \object{HD~168592} &          126.6             &     39.3		  & 1.18 \\
	14 & 54~905.40 & 0.76   & X~Sgr						 & 			126.4			 &	    49.2		& 1.06 \\
	15 & 54~905.41 &			 & \object{HD~169916} &   		 122.6			   &	  45.7		  & 1.11 \\
	16 & 54~905.43 & 0.76	 & X~Sgr						 & 			129.4			 &	    55.3		& 1.02 \\
	\hline
\end{tabular}
\tablefoot{MJD is the modified Julian date, $\phi$ is the pulsation phase at the time of the observations, $B_\mathrm{p}$ is the projected baseline length, PA is the position angle of the projected baseline (from east to north), and AM is the airmass.}
\label{table__journal}
\end{table}

\subsection{Data reduction}

To reduce these data we used two different reduction packages, MIA and EWS\footnote{The MIA+EWS software package is available at http://www.strw.leidenuniv.nl/$\sim$nevec/MIDI/index.html.}. MIA, developed at the Max-Planck-Institut f$\mathrm{\ddot{u}}$r Astronomie, implements an incoherent method where the power spectral density function of each scan is integrated to obtain the squared visibility amplitudes, which are then integrated over time. EWS, developed at the Leiden Observatory, is a coherent analysis that first aligns the interferograms before co-adding them, which result in a better signal-to-noise ratio of the visibility amplitudes. The data reduction results obtained with the MIA and EWS packages agree well within the uncertainties.

The choice of the detector mask for extracting the source spectrum and estimating the background can be critical for the data quality. The latest version of the software uses adaptive masks, where shifts in positions and the width of the mask can be adjusted by fitting the mask for each target. To achieve the best data quality, we first used MIA to fit a specific mask for each target (also allowing a visual check of the data and the mask), and then applied it in the EWS reduction.

Photometric templates from \citet{Cohen_1999_04_0} were employed to perform an absolute calibration of the flux density. We finally averaged the data for a given target. This is justified because the MIDI uncertainties are on the order of 7-15\,\% \citep{Chesneau_2007_10_0}, and the projected baseline and PA are not significantly different for separate observing dates. The uncertainties of the visibilities are mainly dominated by the photometric calibration errors, which are common to all spectral channels ; we accordingly chose the standard deviation over a $1\,\mu$m range as error bars.

\subsection{Flux and visibility fluctuations between datasets}
\label{subsection__flux_and_visibility_fluctuations_between_datasets}

MIDI is strongly sensitive to the atmospheric conditions and can provide  mis-estimates of the thermal flux density and visibility. This can be even worse for datasets combined from different observing nights, for instance for T~Mon in our case. Another source of variance between different datasets can appear from the calibration process, that is, from a poor absolute flux and visibility calibration. In our case, each Cepheid observation was calibrated with a different calibrator (i.e., \#3-4 and \#11-12 for T Mon and \#13-14 and \#15-16 for X Sgr), which enabled us to check the calibrated data.

To quantify the fluctuations, we estimated the spectral relative variation for the flux density and visibility, that is, the ratio of the standard deviation to the mean value for each wavelength between two different calibrated observations. For X~Sgr, the average variation (over all $\lambda$) is lower than 5\,\% on the spectral flux and lower than 1.5\,\% on the visibility. This is a sightly higher for T~Mon because the data were acquired on separate nights ; we measured an average variation lower than 8\,\% on the spectral flux and lower than 4\,\% on the visibility.

\section{Circumstellar envelope modeling}
\label{section__cse_modeling}

\subsection{Visibility and spectral energy distribution}
\label{subsection__visibility_and_sed}

The averaged calibrated visibility and spectral energy distribution (SED) are shown with blue dots in Figs.~\ref{graph__visibility_xsgr} and \ref{graph__visibility_tmon}. The quality of the data in the window $9.4 < \lambda < 10\,\mu$m deteriorates significantly because of the water and ozone absorption in the Earth's atmosphere. Wavelengths longer than $12\,\mathrm{\mu m}$ were not used because of low sensitivity. We therefore only used the spectra outside these wavelengths.

The photosphere of the stars is considered to be unresolved by the interferometer ($V > 98\,\%$), therefore the visibility profile is expected to be equal to unity for all wavelengths. However, we noticed a decreasing profile for both stars. This behavior is typical of emission from a circumstellar envelope (or disk), where the size of the emitting region grows with wavelength. This effect can be interpreted as emission at longer wavelengths coming from cooler material that is located at larger distances from the Cepheid than the warmer material emitted at shorter wavelengths. \citet{Kervella_2009_05_0} previously observed the same trend for $\ell$~Car and RS~Pup.

Assuming that the CSE is resolved by MIDI, the flux contribution of the dust shell is estimated to be about 50\,\% at $10.5\,\mathrm{\mu m}$ for T~Mon and 7\,\% for X~Sgr. It is worth mentioning that the excess is significantly higher for the longer-period Cepheid, T~Mon, adding additional evidence about the correlation between the pulsation period and the CSE brightness suspected previously \citep{Gallenne_2011_11_0,Merand_2007_08_0}.
%V = 0.93 pour X Sgr et 0.66 pout T Mon

The CSE is also detected in the SED, with a contribution progressively increasing with wavelength. Compared with Kurucz atmosphere models \citep[][solid black curve in Fig.~\ref{graph__visibility_xsgr} and \ref{graph__visibility_tmon}]{Castelli_2003__0}, we notice that the CSE contribution becomes significant around $8\,\mathrm{\mu m}$ for X~Sgr, while for T~Mon it seems to start at shorter wavelengths. The Kurucz models were interpolated at $T_\mathrm{eff} = 5900$\,K , $\log g = 2$ and $V_\mathrm{t} = 4\,\mathrm{km~s^{-1}}$ for X~Sgr \citep{Usenko_2012__0}. For T~Mon observed at two different pulsation phases, the stellar temperature only varies from $\sim 5050$\,K ($\phi = 0.33$) to $\sim 5450$\,K ($\phi = 0.12$), we therefore chose the stellar parameters $T_\mathrm{eff} = 5200$\,K , $\log g = 1$, and $V_\mathrm{t} = 4\,\mathrm{km~s^{-1}}$ \citep{Kovtyukh_2005_01_0} for an average phase of 0.22. This has an effect of a few percent in the following fitted parameters (see Sect.~\ref{subsubsection__tmon}).

Given the limited amount of data and the lack of feature that could be easily identified (apart from the alumina shoulder, see below), the investigation of the dust content and the dust grains geometrical properties is therefore limited by the high level of degeneracy. We restricted ourself to the range of dust compound to the refractory ones or the most frequently encountered around evolved stars.

The wind launched by Cepheids is not supposed to be enriched compared with the native composition of the star. Therefore, the formation of carbon grains in the vicinity of these stars is highly unprobable. The polycyclic aromatic hydrocarbons (PAHs) detected around some Cepheids by Spitzer/IRAC and MIPS have an interstellar origin and result from a density enhancement at the interface between the wind and the interstellar medium that leads to a bow shock \citep{Marengo_2010_12_0}. It is noteworthy that no signature of PAHs is observed in the MIDI spectrum or the MIDI visibilities (see Fig.~\ref{graph__visibility_xsgr} and \ref{graph__visibility_tmon}).

The sublimation temperature of iron is higher than that of alumina and rapidly increases with density. Hence, iron is the most likely dust species expected to form in dense (shocked) regions with temperatures higher 1500 K \citep{Pollack_1994_02_0}. Moreover, alumina has a high sublimation temperature in the range of 1200-2000\,K (depending of the local density), and its presence is generally inferred by a shoulder of emission between 10 and $15\,\mathrm{\mu m}$ \citep{Chesneau_2005_06_0,Verhoelst_2009_04_0}. Such a shoulder is identified in the spectrum and visibility of X~Sgr, suggesting that this compound is definitely present. Yet, it must be kept in mind that the low aluminum abundance at solar metallicity prevents the formation of a large amount of this type of dust.  No marked shoulder is observed in the spectrum and visibilities from T Mon, which is indicative of a lower content.  The silicates are easily identified owing to their signature at $10\,\mathrm{\mu m}$. This signature is not clearly detected in the MIDI data.

\subsection{Radiative transfer code: \texttt{DUSTY}}

To model the thermal-IR SED and visibility, we performed radiative transfer calculations for a spherical dust shell. We used the public-domain simulation code \texttt{DUSTY} \citep{Ivezic_1997_06_0,Ivezic_1999_11_0}, which solves the radiative transfer problem in a circumstellar dusty environment by analytically integrating the radiative-transfer equation in planar or spherical geometries. The method is based on a self-consistent equation for the spectral energy density, including dust scattering, absorption, and emission. To solve the radiative transfer problem, the following parameters for the central source and the dusty region are required:
\begin{itemize}
\item the spectral shape of the central source's radiation,
\item the dust grain properties: chemical composition, grain size distribution, and dust temperature at the inner radius,
\item the density distribution of the dust and the relative thickness, and
\item the radial optical depth at a reference wavelength.
\end{itemize}

\texttt{DUSTY} then provides the SED, the surface brightness at specified wavelengths, the radial profiles of density, optical depth and dust temperature, and the visibility profile as a function of the spatial frequency for the specified wavelengths.

\subsection{Single dust shell model}
\label{subsection__single_dust_shell_model}
\begin{figure}[!ht]
\centering
\resizebox{\hsize}{!}{\includegraphics{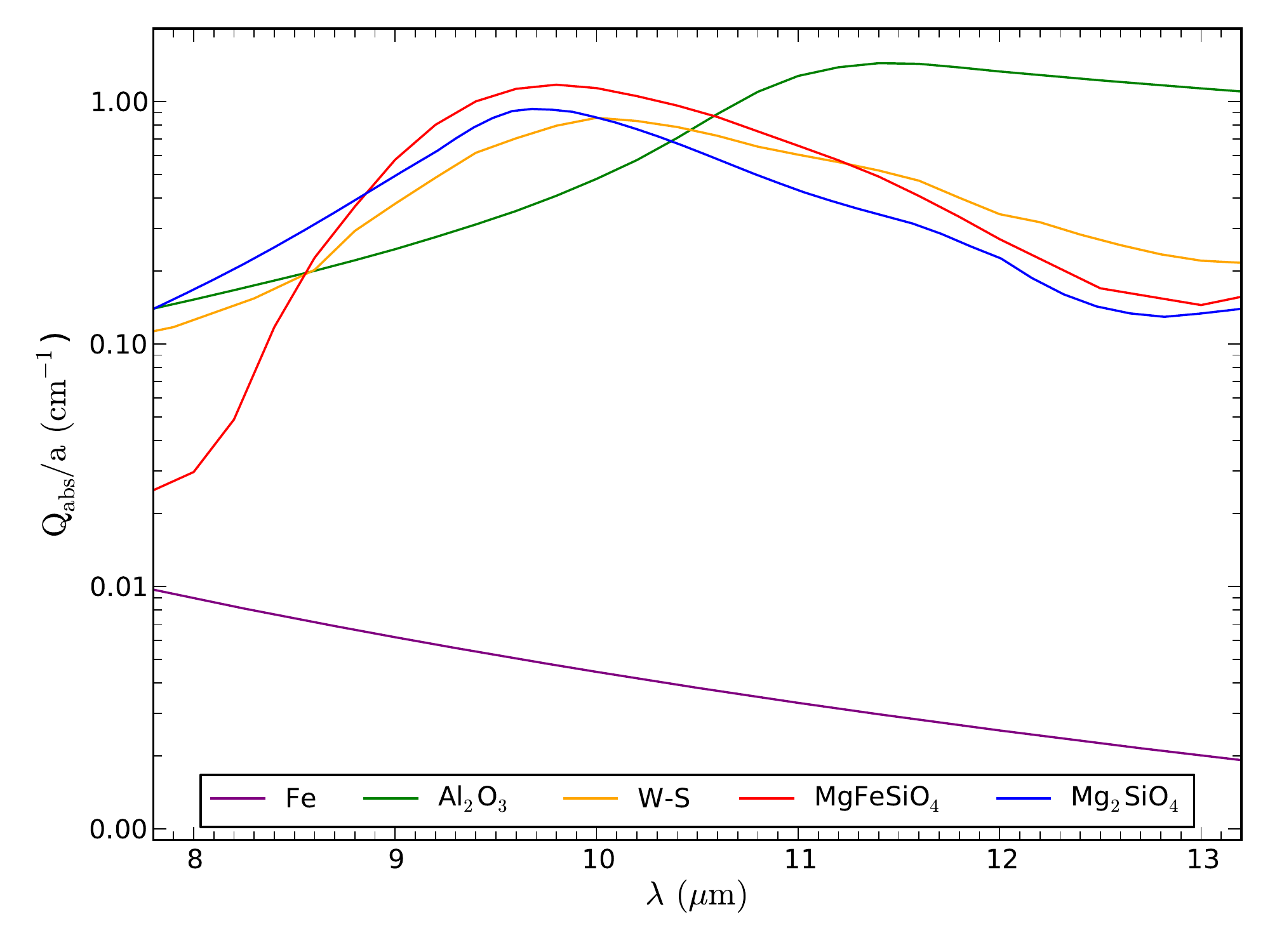}}
\caption{Optical efficiency per grain radius in the MIDI wavelength region for some dust species.}
\label{graph__dust_efficiency}
\end{figure}

We performed a simultaneous fit of the MIDI spectrum and visibilities with various \texttt{DUSTY} models to check the consistency with our data. The central source was represented with Kurucz atmosphere models \citep{Castelli_2003__0} with the stellar parameters listed in Sect.~\ref{subsection__visibility_and_sed}. In the absence of strong dust features, we focused on typical dust species encountered in circumstellar envelopes and according to the typical abundances of Cepheid atmospheres, that is, amorphous alumina \citep[Al$_2$O$_3$ compact,][]{Begemann_1997_02_0}, iron \citep[Fe,][]{Henning_1995_07_0}, warm silicate \citep[W-S,][]{Ossenkopf_1994_11_0}, olivine \citep[MgFeSiO$_4$,][]{Dorschner_1995_08_0}, and forsterite \citep[Mg$_2$SiO$_4$,][]{Jager_2003_09_0}. We present in Fig.~\ref{graph__dust_efficiency} the optical efficiency of these species for the MIDI wavelength region. We see in this plot for instance that the amorphous alumina is optically more efficient around $11\,\mathrm{\mu m}$. We also notice that forsterite, olivine, and warm silicate have a similar optical efficiency, but as we cannot differentiate these dust species with our data, we decided to use warm silicates only.

We used a grain size distribution following a standard Mathis-Rumpl-Nordsieck (MRN) relation \citep{Mathis_1977_10_0}, that is, $n(a) \propto a^{3.5}$ for $0.005 \leqslant a \leqslant 0.25\,\mathrm{\mu m}$. We chose a spherical density distribution in the shell following a radiatively driven wind, because Cepheids are giant stars and might lose mass via stellar winds \citep{Neilson_2008_09_0}. In this case, \texttt{DUSTY} computes the density structure by solving the hydrodynamics equations, coupled to the radiative transfer equations. The shell thickness is the only input parameter required. It is worth mentioning that we do not know the dust density profile in the Cepheid outflow, and we chose the hydrodynamic calculation in DUSTY as a good assumption.

For both stars, we also added $B, V, J, H$ and $K$ photometric light curves from the literature to our mid-IR data to better constrain the stellar parameters (\citealt{Moffett_1984_07_0,Berdnikov_2008_04_0, Feast_2008_06_0} for X~Sgr, \citealt{Moffett_1984_07_0,Coulson_1985__0,Berdnikov_2008_04_0,Laney_1992_04_0} for T~Mon). To avoid phase mismatch, the curves were fitted with a cubic spline function and were interpolated at our pulsation phase. We then used these values in the fitting process. The conversion from magnitude to flux takes into account the photometric system and the filter bandpass. During the fitting procedure, all flux densities $< 3\,\mathrm{\mu m}$ were corrected for interstellar extinction $A_\lambda = R_\lambda E(B - V)$ using the total-to-selective absorption ratios $R_\lambda$ from \citet{Fouque_2003__0} and \citet{Hindsley_1989_06_0}. The mid-IR data were not corrected for the interstellar extinction, which we assumed to be negligible.

The free parameters are the stellar luminosity ($L_\star$), the dust temperature at the inner radius ($T_\mathrm{in}$), the optical depth at $0.55\,\mathrm{\mu m}$ ($\tau_\mathrm{0.55\mu m}$), and the color excess $E(B - V)$. Then we extracted from the output files of the best-fitted \texttt{DUSTY} model the shell internal diameter ($\theta_\mathrm{in}$), the stellar diameter ($\theta_\mathrm{LD}$), and the mass-loss rate $\dot{M}$. The stellar temperature of the Kurucz model ($T_\mathrm{eff}$), the shell's relative thickness and the dust abundances were fixed during the fit. We chose $R_\mathrm{out}/R_\mathrm{in} = 500$ for the relative thickness as it is not constrained with our mid-IR data. The distance of the star was also fixed to 333.3\,pc for X~Sgr \citep{Benedict_2007_04_0} and 1309.2\,pc for T~Mon \citep{Storm_2011_10_0}. 

\subsection{Results}

\begin{figure}[!ht]
\centering
\resizebox{\hsize}{!}{\includegraphics{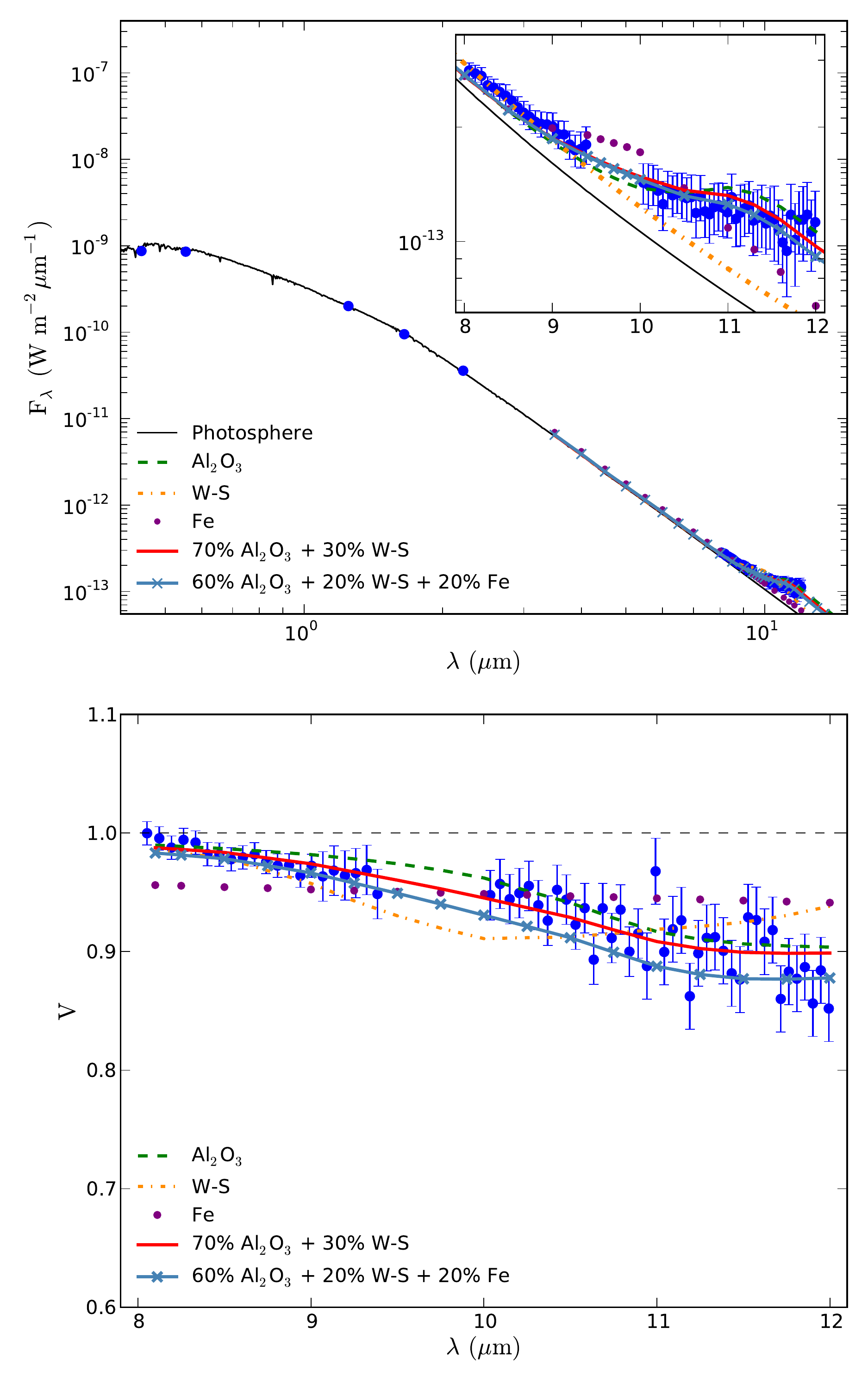}}
\caption{Calibrated visibility and spectrum of X~Sgr (blue dots with error bars). The solid black line in the upper panel represents the photosphere of the Cepheid modeled with Kurucz's spectra, while the dashed black line in the lower panel stands for an unresolved star for comparison. Other color curves are the fitted models.}
\label{graph__visibility_xsgr}
\end{figure}
\begin{figure}[!ht]
\centering
\resizebox{\hsize}{!}{\includegraphics{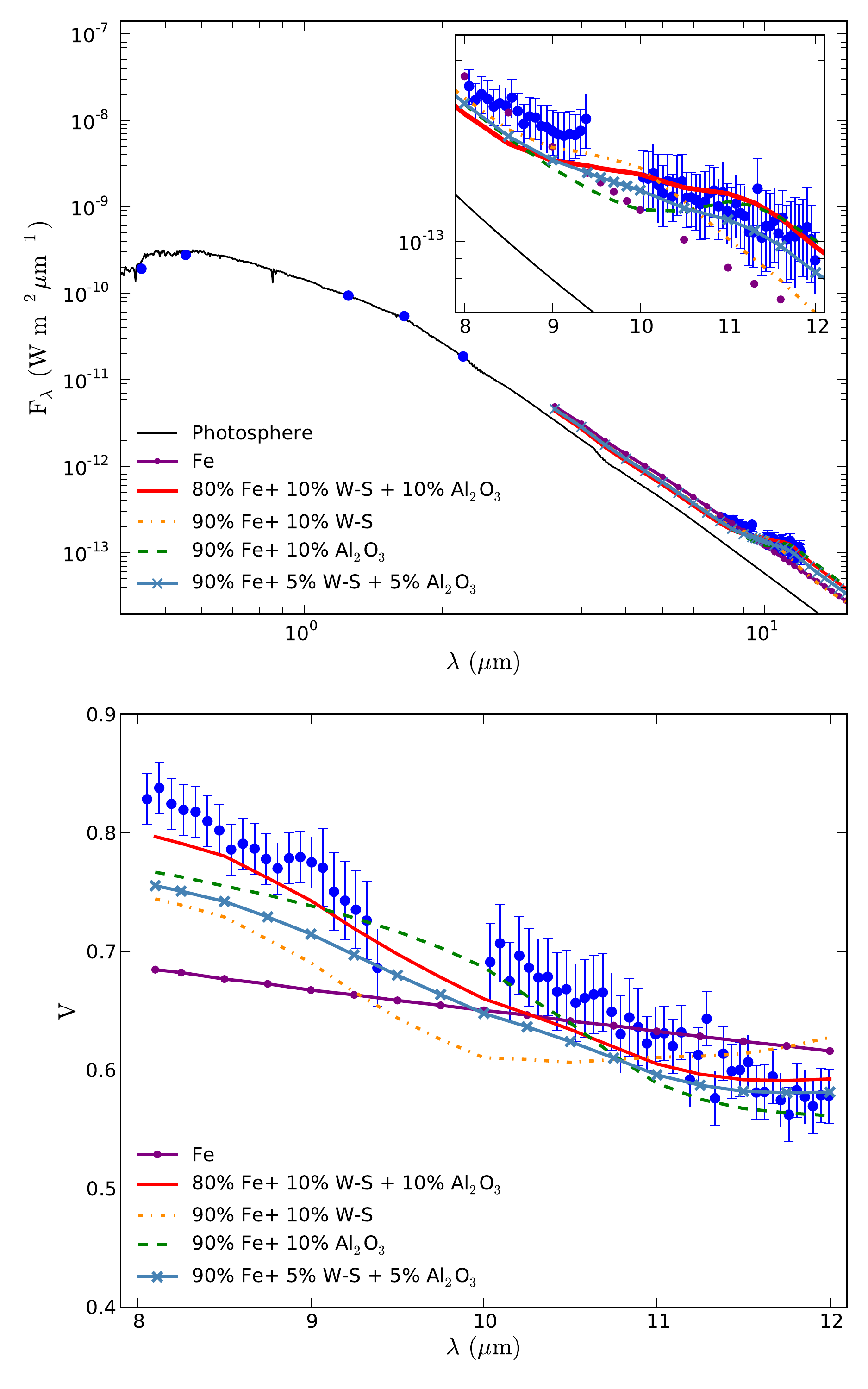}}
\caption{Calibrated visibility and spectrum of T~Mon (blue dots with error bars). The solid black line in the upper panel represents the photosphere of the Cepheid modeled with Kurucz's spectra. Other color curves are the fitted models.}
\label{graph__visibility_tmon}
\end{figure}

\subsubsection{X~Sgr} 

The increase of the SED around $11\,\mathrm{\mu m}$ made us investigate in the direction of a CSE composed of Al$_2$O$_3$ material, which is optically efficient at this wavelength. After trying several dust species, we finally found a good agreement with a CSE composed of 100\,\% amorphous alumina (model \#1 in Table~\ref{table__fit_result}). The fitted parameters are listed in Table~\ref{table__fit_result} and are plotted in Fig.~\ref{graph__visibility_xsgr}. 
However, a dust composed of 70\,\% Al$_2$O$_3$ + 30\,\% W-S (model \#4), or dust including some iron (model \#5), are also statistically consistent with our observations. Consequently, we chose to take as final parameters and uncertainties the average values and standard deviations (including their own statistical errors added quadratically) between models \#1, \#4, and \#5. The final adopted parameters are listed in Table~\ref{table__fit_results_final}. It is worth mentioning that for these models all parameters have the same order of magnitude. The error on the stellar angular diameter was estimated from the luminosity and distance uncertainties.

The CSE of X~Sgr is optically thin ($\tau_\mathrm{0.55\mu m} = 0.0079 \pm 0.0021$) and has an internal shell diameter of $\theta_\mathrm{in} = 15.6 \pm 2.9$\,mas. The condensation temperature we found is in the range of what is expected for this dust composition (1200-1900\,K). The stellar angular diameter (and in turn the luminosity) is also consistent with the value estimated from the surface-brightness method at that pulsation phase \citep[][$1.34 \pm 0.03$\,mas]{Storm_2011_10_0} and agrees with the average diameter measured by \citet[][$1.47 \pm 0.03\,\mathrm{\mu m}$]{Kervella_2004_03_0}. The relative CSE excess in the VISIR PAH1 filter of $13.3 \pm 0.5$\,\% also agrees with the one estimated by \citet[][$11.7 \pm 4.7\,\%$]{Gallenne_2011_11_0}. Our derived color excess $E(B-V)$ is within $1\sigma$ of the average value $0.227 \pm 0.013$ estimated from photometric, spectroscopic, and space reddenings \citep{Fouque_2007_12_0,Benedict_2007_04_0,Kovtyukh_2008_09_0}.

\subsubsection{T~Mon}
\label{subsubsection__tmon}

The CSE around this Cepheid has a stronger contribution than X~Sgr. The large excess around $8\,\mathrm{\mu m}$ enables us to exclude a CSE composed of 100\,\% Al$_2$O$_3$, because of its low efficiency in this wavelength range. We first considered dust composed of iron. However, other species probably contribute to the opacity enhancement. As showed in Fig.~\ref{graph__visibility_tmon}, a 100\,\% Fe dust composition is not consistent with our observations. We therefore used a mixture of W-S, Al$_2$O$_3$ and Fe to take into account the optical efficiency at all wavelengths. The best model that agrees with the visibility profile and the SED is model \#5, including 90\,\% Fe + 5\,\% Al$_2$O$_3$ + 5\,\% W-S. The fitted parameters are listed in Table~\ref{table__fit_result} and are plotted in Fig.~\ref{graph__visibility_tmon}. However, because no specific dust features are present to constrain the models, other dust compositions are also consistent with the observations. Therefore we have chosen the average values and standard deviations (including their own statistical errors added quadratically) between models \#2, \#4 and \#5 as final parameters and uncertainties. The final adopted parameters are listed in Table~\ref{table__fit_results_final}. 

The choice of a stellar temperature at $\phi = 0.33$ or 0.12 in the fitting procedure (instead of an average pulsation phase as cited in Sect.~\ref{subsection__visibility_and_sed}) changes the derived parameters by at most 10\,\% (the variation of the temperature is lower in the mid-IR). To be conservative, we added quadratically this relative error to all parameters of Table~\ref{table__fit_results_final}.

The CSE of T~Mon appears to be thicker than that of X~Sgr, with ($\tau_\mathrm{0.55\mu m} = 0.151 \pm 0.042$), and an internal shell diameter of $\theta_\mathrm{in} = 15.9 \pm 1.7$\,mas. The derived stellar diameter agrees well with the $1.01 \pm 0.03$\,mas estimated by \citet[][at $\phi = 0.22$]{Storm_2011_10_0}. The deduced color excess $E(B-V)$ agrees within $1\sigma$ with the average value $0.181 \pm 0.010$ estimated from photometric, spectroscopic and space reddenings \citep{Fouque_2007_12_0,Benedict_2007_04_0,Kovtyukh_2008_09_0}.
We derived a particularly high IR excess in the VISIR PAH1 filter of $87.8 \pm 9.9$\,\%, which might make this Cepheid a special case. It is worth mentioning that we were at the sensitivity limit of MIDI for this Cepheid, and the flux might be biased by a poor subtraction of the thermal sky background. However, the clear decreasing trend in the visibility profile as a function of wavelength cannot be attributed to a background emission, and we argue that this is the signature of a CSE. In Sect.~ \ref{section__discussion} we make a comparative study to remove the thermal sky background and qualitatively estimate the unbiased IR excess.

\begin{sidewaystable*}[]
\centering
\caption{The fit model parameters.}
\begin{tabular}{ccccccccccccc} 
\hline
\hline
 Model		&	$L_\star$		& 	$T_\mathrm{eff}$\tablefootmark{d} & $\theta_\mathrm{LD}$ & $E(B - V)$ &	$T_\mathrm{in}$ &  $\theta_\mathrm{in}$	&	$\tau_\mathrm{0.55\mu m}$			& $\dot{M}$	& $\alpha$  & $\chi^2_\mathrm{r}$ & \# \\
			&	$(L_\odot)$	   &			(K)				   						& (mas)						& & (K)	&	(mas)							&		($\times10^{-3}$)				&	($M_\odot\,yr^{-1}$)	& (\%) & &  \\
\hline		
\multicolumn{12}{c}{\textbf{X~Sgr}}   & \\
Al$_2$O$_3$\tablefootmark{a}	& 	$2151 \pm 34$ 	&  5900	& $1.24 \pm 0.08$ & $0.199 \pm 0.009$ & $1732 \pm 152$	&  13.2 	&	$6.5 \pm 0.8$	&	$5.1\times10^{-8}$	& 13.0 & 0.78 & 1 \\

W-S\tablefootmark{b} &	$2155 \pm 40$	&	5900 &$1.24 \pm 0.08$ & $0.200 \pm 0.011$ & $1831 \pm 141$	&	14.7	&	$11.8 \pm 0.2$	&	$6.4\times10^{-8}$			& 15.0 & 1.53 & 2 \\

Fe\tablefootmark{c} 	&	$2306 \pm 154$ &	5900 & $1.28 \pm 0.09$ & $0.230 \pm 0.031$ & $1456 \pm 605$	&	29.3	&	$8.9 \pm 4.4$	&	$6.2\times10^{-8}$	& 9.5 & 4.30 & 3\\

70\,\% Al$_2$O$_3$ + 30\,\% W-S	&  $2153 \pm 31$ & 5900	& $1.24 \pm 0.08$ & $0.200 \pm 0.009$ & $1519 \pm 117$ & 19.6 & $6.6 \pm 0.7$ & $5.6\times10^{-8}$ &  13. 7  & 0.58 &  4\\

60\,\% Al$_2$O$_3$ + 20\,\% W-S + 20\,\% Fe & $2160 \pm 35$ & 5900 & $1.24 \pm 0.08$ & $0.201 \pm 0.009$ & $1802 \pm 130$ & 13.9 & $10.6 \pm 1.2$ & $6.1\times10^{-8}$ & 13.3 & 0.68 & 5\\

\hline		
\multicolumn{12}{c}{\textbf{T~Mon}} & \\

Fe 	&	$12~453 \pm 775$			&	5200	& $0.98 \pm 0.03$ & $0.183 \pm 0.049$ & $1190 \pm 59$	&	29.4	&	$113 \pm 21$	&	$5.0\times10^{-7}$	& 99.5 & 5.36 & 1 \\

80\,\% Fe + 10\,\% W-S + 10\,\% Al$_2$O$_3$	& $11~606 \pm 434$ & 5200 &  $0.94 \pm 0.03$ &  $0.144 \pm 0.028$ &  $1418 \pm 42$	& 16.6	& $126 \pm 13$ & $4.4\times10^{-7}$ & 82.2  & 1.52 & 2\\

 90\,\% Fe + 10\,\% W-S & $11~696 \pm 667$ &  5200 	& $0.95 \pm 0.03$ & $0.149 \pm 0.044$ &  $1389 \pm 54$	& 18.0	&  $147 \pm 23$	&  $5.0\times10^{-7}$	&   94.6  &   3.04 &   3\\
 
 90\,\% Fe + 10\,\% Al$_2$O$_3$ & $11~455 \pm 580$ & 5200 &  $0.94 \pm 0.03$ &  $0.137 \pm 0.040$ &  $1439 \pm 49$	& 15.9	&  $158 \pm 21$	&  $5.0\times10^{-7}$	&   87.5  &   2.21 &   4\\
 
 90\,\% Fe + 5\,\% Al$_2$O$_3$ + 5\,\% W-S & $11~278 \pm 597$ &	5200 &  $0.93 \pm 0.04$ & $0.125 \pm 0.042$ & $1458 \pm 48$	& 15.3 & $170 \pm 23$ & $5.2\times10^{-7}$ & 93.6 & 2.22 & 5\\
 
\hline 
\end{tabular}
\tablefoot{$L_\star$, $T_\mathrm{eff}$ and $\theta_\mathrm{LD}$ are the stellar luminosity,  temperature and angular diameter, respectively. $T_\mathrm{in}$ and $\theta_\mathrm{in}$ are respectively the dust temperature at the inner radius and the inner diameter of the dust shell. $\tau_\mathrm{0.55\mu m}$ is the optical depth at $0.55\,\mathrm{\mu m}$ and $\dot{M}$ denotes the mass-loss rate. $\alpha$ stands for the relative excess (ratio of the envelope to the photosphere flux). 
\tablefoottext{a}{from \citet{Begemann_1997_02_0}.}
\tablefoottext{b}{Warm silicate from \citet{Ossenkopf_1994_11_0}.}
\tablefoottext{c}{Iron from \citet{Henning_1996_07_0}.}
\tablefoottext{d}{Hold fixed.}
}
\label{table__fit_result}
\end{sidewaystable*}

\begin{table}[!h]
\centering
\caption{Final adopted parameters.}
\begin{tabular}{ccccccccc} 
 \hline
 \hline
 	&	X~Sgr	&	T~Mon \\
 \hline
	$L_\star$ $(L_\odot)$													&	$2155 \pm 58$		&	$11~446 \pm 1486$		\\
	$T_\mathrm{eff}$	(K)													&	5900							&	5200							\\
	$\theta_\mathrm{LD}$ (mas)										&	$1.24 \pm 0.14$			&	$0.94 \pm 0.11$		\\
	$E(B - V)$																		&	$0.200 \pm 0.032$	&	$0.135 \pm 0.066$	\\
	$T_\mathrm{in}$	(K)														&	$1684 \pm 225$			&	$1438 \pm 166$		\\
	$\theta_\mathrm{in}$ (mas)										&	$15.6 \pm 2.9$ 			&	 $15.9 \pm 1.7$ 		\\
	$\tau_\mathrm{0.55\mu m}$ 	($\times10^{-3}$)	&	$7.9 \pm 2.1$			&	$151 \pm 42$			\\
	$\dot{M}$ ($\times10^{-8} M_\odot\,yr^{-1}$)		&	$5.6 \pm 0.6$			&	$48.7 \pm 5.9$			\\
	$\alpha$ (\%)																&	$13.3 \pm 0.7$			&	$87.8 \pm 9.9$\tablefootmark{*}\\
 \hline
\end{tabular}
\tablefoot{Averaged parameters from the fitted models. See Sect.~\ref{section__cse_modeling} for more details. \tablefoottext{*}{This value is likely to be biased by the sky background (see Sect.~\ref{section__discussion}).} }
\label{table__fit_results_final}
\end{table}

\section{Period-excess relation}
\label{section__period_excess_relation}

\begin{figure}[!ht]
\centering
\resizebox{\hsize}{!}{\includegraphics{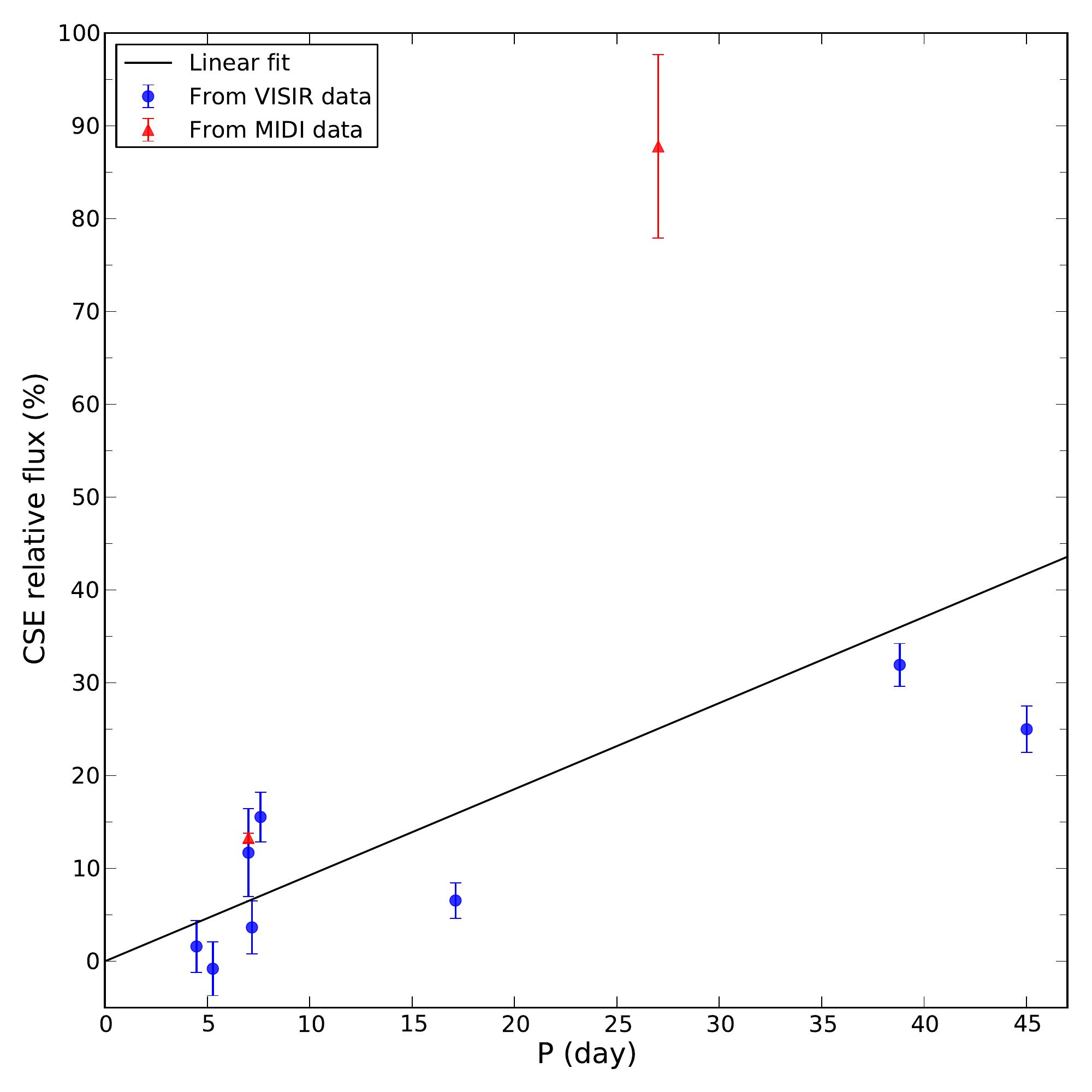}}
\caption{Measured relative CSE fluxes in the VISIR PAH1 filter for nine Cepheids as a function of the pulsation period.}
\label{graph__excess}
\end{figure}

\citet{Gallenne_2011_11_0} presented a probable correlation between the pulsation period and the CSE relative excess in the VISIR PAH1 filter. From our fitted \texttt{DUSTY} model, we estimated the CSE relative excess by integrating over the PAH1 filter profile. This allowed us another point of view on the trend of this correlation. X~Sgr was part of the sample of \citet{Gallenne_2011_11_0} and can be directly compared with our result, while T~Mon is a new case. This correlation is plotted in Fig.~\ref{graph__excess}, with the measurements of this work as red triangles. The IR excess for X~Sgr agrees very well with our previous measurements \citep{Gallenne_2011_11_0}. The excess for T~Mon is extremely high, and does not seem to follow the suspected linear correlation. 

Fig.~\ref{graph__excess} shows that longer-period Cepheids have higher IR excesses. This excess is probably linked to past or ongoing mass-loss phenomena. Consequently, this correlation shows that long-period Cepheids have a larger mass-loss than shorter-period, less massive stars. This behavior might be explained by the stronger velocity fields in longer-period Cepheids, and the presence of shock waves at certain pulsation phases \citep{Nardetto_2006_07_0,Nardetto_2008_10_0}. This scenario is consistent with the theoretically predicted range, $10^{-10}$--$10^{-7} M_\odot\,yr^{-1}$, of \citet{Neilson_2008_09_0}, based on a pulsation-driven mass-loss model. \citet{Neilson_2011_05_0} also found that a pulsation-driven mass-loss model combined with moderate convective-core overshooting provides an explanation for the Cepheid mass discrepancy, where stellar evolution masses differ by 10-20\,\% from stellar pulsation calculations.

We fitted the measured mi-IR excess with a linear function of the form
\begin{displaymath}
 f_\mathrm{8.6\,\mu m} = \alpha_\mathrm{8.6\,\mu m}P,
 \end{displaymath}
with $f$ in \% and $P$ in day. We used a general weighted least-squares minimization, using errors on each measurements as weights . We found a slope $\alpha_\mathrm{8.6\,\mu m} = 0.83 \pm 0.04\,\mathrm{\%.d^{-1}}$, including T~Mon, and $\alpha_\mathrm{8.6\,\mu m} = 0.81 \pm 0.04\,\mathrm{\%.d^{-1}}$ without. The linear relation is plotted in Fig.~\ref{graph__excess}.

\section{Discussion}
\label{section__discussion}

Since the first detection around $\ell$~Car \citep{Kervella_2006_03_0}, CSEs have been detected around many other Cepheids \citep{Gallenne_2011_11_0,Merand_2007_08_0,Merand_2006_07_0}. Our works, using IR and mid-IR high angular resolution techniques, lead to the hypothesis that all Cepheids might be surrounded by a CSE. The mechanism for their formation is still unknown, but it is very likely a consequence of mass loss during the pre-Cepheid evolution stage or during the multiple crossings of the instability strip. The period--excess relation favors the last scenario, because long-period Cepheids have higher masses and cross the instability strip up to three times.

Other mid- and far-IR  extended emissions have also been reported by \citet{Barmby_2011_11_0} around a significant fraction of their sample (29 Cepheids), based on Spitzer telescope observations. The case of $\delta$~Cep was extensively discussed in \citet{Marengo_2010_12_0}. From IRAS observations, \citet{Deasy_1988_04_0} also detected IR excesses and estimated mass-loss rate ranging from $10^{-10}$ to $10^{-6}M_\odot\,yr^{-1}$. The values given by our \texttt{DUSTY} models agree. They are also consistent with the predicted mass-loss rate from \citet{Neilson_2008_09_0}, ranging from $10^{-10}$ to $10^{-7}M_\odot\,yr^{-1}$.

These CSEs might have an impact on the Cepheid distance scale through the photometric contribution of the envelopes. While at visible and near-IR wavelengths the CSE flux contribution might be negligible ($< 5$\,\%), this is not the case in the mid-IR domain \citep[see][for a more detailed discussion]{Kervella_2013_02_0}. This is particularly critical because near- and mid-IR P-L relation are preferred due to the diminished impact of dust extinction. Recently, \citet{Majaess_2013_08_0} re-examined the 3.6 and 4.5\,$\mathrm{\mu m}$ Spitzer observations and observed a nonlinear trend on the period-magnitude diagrams for LMC and SMC Cepheids. They found that longer-period Cepheids are slightly brighter than short-period ones. This trend is compatible with our period-excess relation observed for Galactic Cepheids. \citet{Monson_2012_11_0} derived Galactic P--L relations at 3.6 and 4.5\,$\mathrm{\mu m}$ and found a strong color variation for Cepheids with $P > 10$\,days, but they attributed this to enhanced CO absorption at $4.5\,\mathrm{\mu m}$. From their light curves, we estimated the magnitudes expected at our observation phase for X~Sgr and T~Mon \citep[using the ephemeris from][]{Samus_2009_01_0} to check the consistency with the values given by our \texttt{DUSTY} models (integrated over the filter bandpass). For X~Sgr, our models give averaged magnitudes $m_\mathrm{3.6\,\mathrm{\mu m}} = 2.55 \pm 0.06$ and $m_\mathrm{4.5\,\mathrm{\mu m}} = 2.58 \pm 0.05$ (taking into account the 5\,\% flux variations of Sect.~\ref{subsection__flux_and_visibility_fluctuations_between_datasets}), to be compared with $2.54 \pm 0.02$ and $2.52 \pm 0.02$ from \citet{Monson_2012_11_0}. For T~Mon, we have $m_\mathrm{3.6\,\mathrm{\mu m}} = 2.94 \pm 0.14$ and $m_\mathrm{4.5\,\mathrm{\mu m}} = 2.94 \pm 0.14$ from the models (taking into account the 8\,\% flux variations of Sect.~\ref{subsection__flux_and_visibility_fluctuations_between_datasets} and a 10\,\% flux error for the phase mismatch), to be compared with $3.29 \pm 0.08$ and $3.28 \pm 0.05$ (with the rms between phase 0.12 and 0.33 as uncertainty). Our estimated magnitudes are consistent for X~Sgr, while we differ by about $2\sigma$ for T~Mon. As we describe below, we suspect a sky background contamination in the MIDI data. The estimated excesses from the model at 3.6 and $4.5\,\mathrm{\mu m}$ are $6.0 \pm 0.5$\,\% and $6.3 \pm 0.5$\,\% for X~Sgr, and $46 \pm 5$\,\% and $58 \pm 6$\,\% for T~Mon (errors estimated from the standard deviation of each model). This substantial photometric contribution probably affects the Spitzer/IRAC P--L relation derived by \citet{Monson_2012_11_0} and the calibration of the Hubble constant by \citet{Freedman_2012_10_0}.

We also compared our models with the Spitzer 5.8 and 8.0\,$\mathrm{\mu m}$ magnitudes of \citet{Marengo_2010_01_0} which are only available for T~Mon. However, their measurements correspond to the pulsation phase 0.65, so we have to take a phase mismatch into account. According to the light curves of \citet{Monson_2012_11_0}, the maximum amplitude at 3.6 and 4.5\,$\mathrm{\mu m}$ is decreasing from 0.42 to 0.40\,mag, respectively. As the light curve amplitude is decreasing with wavelength, we can safely assume a maximum amplitude at 5.8 and 8.0\,$\mathrm{\mu m}$ of 0.25\,mag. We take this value as the highest uncertainty, which we added quadratically to the measurements of \citet{Marengo_2010_01_0}, which leads to $m_\mathrm{5.8\,\mathrm{\mu m}} = 3.43 \pm 0.25$ and $m_\mathrm{8.0\,\mathrm{\mu m}} = 3.32 \pm 0.25$. Integrating our models on the Spitzer filter profiles, we obtained $m_\mathrm{5.8\,\mathrm{\mu m}} = 2.85 \pm 0.14$ and $m_\mathrm{8.0\,\mathrm{\mu m}} = 2.67 \pm 0.14$, which differ by about $2\sigma = 0.5\,$mag from the empirical values at $8\,\mathrm{\mu m}$. A possible explanation of this discrepancy would be a background contamination in our MIDI measurements. Indeed, due to its faintness, T~Mon is at the sensitivity limit of the instrument, and the sky background can contribute to the measured IR flux (only contributes to the incoherent flux). Assuming that this $2\sigma$ discrepancy is due to the sky background emission, we can estimate the contribution of the CSE with the following approach. The flux measured by \citet{Marengo_2010_01_0} corresponds to $f_\star + f_\mathrm{env}$, that is, the contribution of the star and the CSE, while MIDI measured an additional term corresponding to the background emission, $f_\star + f_\mathrm{env} + f_\mathrm{sky}$. From our derived \texttt{DUSTY} flux ratio (Table~\ref{table__fit_results_final}) and the magnitude difference between MIDI and Spitzer, we have the following equations:
\begin{equation}
\label{eq__1}
\dfrac{f_\mathrm{env} + f_\mathrm{sky}}{f_\star} = \alpha,\ \mathrm{and}
\end{equation}
\begin{equation}
\label{eq__2}
2\sigma = -2.5\,\log \left( \dfrac{f_\star +  f_\mathrm{env}}{f_\star + f_\mathrm{env} + f_\mathrm{sky}} \right),
\end{equation}
where $2\sigma$ is the magnitude difference between the Spitzer and MIDI observations. Combining Eqs.~\ref{eq__1} and \ref{eq__2}, we estimate the real flux ratio to be $f_\mathrm{env}/f_\star \sim 19\,$\%. Interestingly, this is also more consistent with the expected period-excess relation plotted in Fig.~\ref{graph__excess}, although in a different filter.

We also derived the IR excess in the $K$ band to check the possible impact on the usual P--L relation for those two stars. Our models gives a relative excess of $\sim 24.3 \pm 2.7$\,\% for T~Mon, and for X~Sgr we found $4.3 \pm 0.3$\%. However, caution is required with the excess of T~Mon since it might suffer from sky-background contamination. Therefore, we conclude that the bias on the $K$-band P--L relation might be negligible compared with the intrinsic dispersion of the P--L relation itself.

\section{Conclusion}
\label{section__conclusion}

Based on mid-IR observations with the MIDI instrument of the VLTI, we have detected the circumstellar envelope around the Cepheids X~Sgr and T~Mon. We used the numerical radiative transfer code \texttt{DUSTY} to simultaneously fit the SED and visibility profile to determine physical parameters related to the stars and their dust shells. We confirm the previous IR emission detected by \citet{Gallenne_2011_11_0} for X~Sgr with an excess of 13.3\,\%, and we estimate a $\sim 19\,$\% excess for T~Mon at $8\,\mathrm{\mu m}$.

As the investigation of the dust content and the dust grains geometrical properties are limited by a high level of degeneracy, we restricted ourselves to typical dust composition for circumstellar environment. We found optically thin envelopes with an internal dust shell radius in the range 15-20\,mas. The relative CSE excess seems to be significant from $8\,\mathrm{\mu m}$ ($> 10$\,\%), depending on the pulsation period, while for shorter wavelengths, the photometric contribution might be negligible. Therefore, the impact on the $K$-band P--L relation is low ($\lesssim 5$\,\%), but it is considerable for the mid-IR P--L relation \citep{Ngeow_2012_09_0,Monson_2012_11_0}, where the bias due to the presence of a CSE can reach more than 30\,\%. Although still not statistically significant, we derived a linear period-excess relation, showing that longer-period Cepheids exhibit a higher IR excess than shorter-period Cepheids.

It is now necessary to increase the statistical sample and investigate whether CSEs are a global phenomena for Cepheids. Interferometric imaging with the second-generation instrument VLTI/MATISSE \citep{Lopez_2006_07_0} will also be useful for imaging and probing possible asymmetry of these CSEs.

%--------------------ACKNOWLEDGEMENTS--------------------

\begin{acknowledgements}
The authors thank the ESO-Paranal VLTI team for supporting the MIDI observations. We also thank the referee for the comments that helped to improve the quality of this paper. A.G. acknowledges support from FONDECYT grant 3130361. W.G. gratefully acknowledge financial support for this work from the BASAL Centro de Astrof\'isica y Tecnolog\'ias Afines (CATA) PFB-06/2007. This research received the support of PHASE, the high angular resolution partnership between ONERA, Observatoire de Paris, CNRS, and University Denis Diderot Paris 7. This work made use of the SIMBAD and VIZIER astrophysical database from CDS, Strasbourg, France and the bibliographic informations from the NASA Astrophysics Data System.
\end{acknowledgements}

%--------------------BIBLIOGRAPHY--------------------

\bibliographystyle{aa}   % if natbib is available
\bibliography{./bibliographie}

\end{document}